\documentclass[prb,superscriptaddress,twocolumn,preprintnumbers,a4,showpacs,floatfix]{revtex4}
\usepackage{amsmath, amssymb, psfrag, tabls, dcolumn, bm, color}
\usepackage[sort&compress]{natbib}
\usepackage[pdftex]{graphicx}
\usepackage{graphicx}

\newcommand\smallurl[1]{{\tiny \url{#1}}}

\newcommand{\be}{\begin{equation}}
\newcommand{\ee}{\end{equation}}
\newcommand{\bea}{\begin{equation*}}
\newcommand{\eea}{\end{equation*}}
\newcommand{\ba}{\begin{array}}
\newcommand{\ea}{\end{array}}
\newcommand{\beqa}{\begin{eqnarray}}
\newcommand{\eeqa}{\end{eqnarray}}
\newcommand{\beqaa}{\begin{eqnarray*}}
\newcommand{\eeqaa}{\end{eqnarray*}}

\newcommand{\matr}{\left( \begin{array}}
\newcommand{\ematr}{\end{array} \right)}

\newcommand{\der}{{\rm d}}

\newcommand{\lsim}{{\;\raise0.3ex\hbox{$<$\kern-0.75em\raise-1.1ex\hbox{$\sim$}}
\;}}
\newcommand{\gsim}{{\;\raise0.3ex\hbox{$>$\kern-0.75em\raise-1.1ex\hbox{$\sim$}}
\;}}

\usepackage{mdwlist}

\def\bcols{\begin{columns}}
\def\ecols{\end{columns}}
\def\bcol{\begin{column}}
\def\ecol{\end{column}}
\def\bit{\begin{itemize}}
\def\eit{\end{itemize}}
\def\bitt{\begin{itemize*}}
\def\eitt{\end{itemize*}}
\def\ben#1{\begin{enumerate}[#1]}
\def\een{\end{enumerate}}
\def\colb#1{\begin{columns}\begin{column}{#1}}

\def\cole{\end{column}\end{columns}}

\usepackage{color}
\usepackage{framed}
\usepackage{comment}
\definecolor{shadecolor}{gray}{0.875}
\specialcomment{answer}{\begin{shaded}}{\end{shaded}}
\specialcomment{vastaus}{}{}
\specialcomment{pikavastaus}{}{}

\RequirePackage{ifthen}
\newboolean{isvastaus}
\setboolean{isvastaus}{false}

\ifthenelse{\boolean{isvastaus}}
{}
{\excludecomment{vastaus}
}

\usepackage{bm}

\begin{document}

\title{Graphene nanoribbons subject to gentle bends}

\author{P. Koskinen}
\email[email:]{pekka.koskinen@iki.fi}
\affiliation{NanoScience Center, Department of Physics, University of Jyv\"askyl\"a, 40014 Jyv\"askyl\"a, Finland}




\begin{abstract}
Since graphene nanoribbons are thin and flimsy, they need support. Support gives firm ground for applications, and adhesion holds ribbons flat, although not necessarily straight: ribbons with high aspect ratio are prone to bend. The effects of bending on ribbons' electronic properties, however, are unknown. Therefore, this article examines the electromechanics of planar and gently bent graphene nanoribbons. Simulations with density-functional tight-binding and revised periodic boundary conditions show that gentle bends in armchair ribbons can cause significant widening or narrowing of energy gaps. Moreover, in zigzag ribbons sizeable energy gaps can be opened due to axial symmetry breaking, even without magnetism. These results infer that, in the electronic measurements of supported ribbons, such bends must be heeded.
\end{abstract}

\maketitle

\section{Introduction}

Graphene nanoribbons (GNRs) are atomically thin and only nanometers wide, which makes them the flimsiest materials in the world. Today such ribbons, acclaimed for promising applications, are fabricated in many ways\cite{kosynkin_nature_09,cai_nature_10}, and investigated for heat conduction\cite{wei_NT_11}, edge features\cite{jia_science_09,suenaga_nature_10,acik_JJAP_11}, and electronic characteristics\cite{wang_nnano_11}, among many other properties. However, since the ribbons are flimsy, they need stabilizing support---although even then ribbons can get folded, torn, rippled, and bent.\cite{wang_nnano_11,li_science_08,jiao_nature_09,jiao_nnano_10}

\begin{figure}[b!]
\includegraphics[width=8cm]{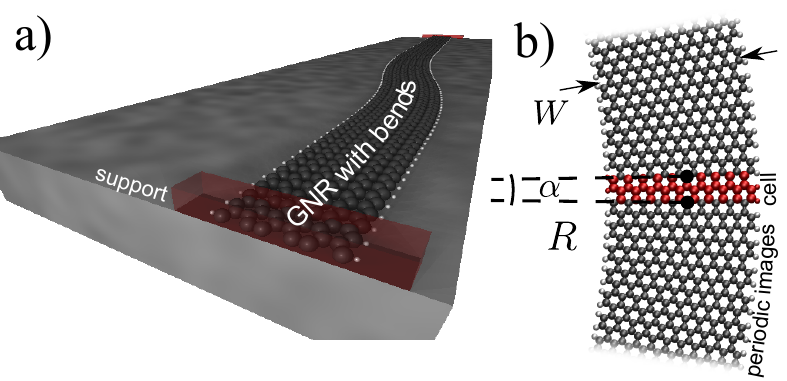}
\caption{(Color online) (a) Supported graphene nanoribbon sketched for gentle, planar bends. (b) The red (dark gray) atoms constitute the unit cell, simulated with revised periodic boundary conditions; the symmetry operation is rotation of an angle $\alpha$ around given origin (not shown). Ribbon's width $W$ is defined by the outmost carbon atoms and $R$ is the mean radius of curvature. The bending parameter of Eq.(\ref{eq:theta}), here $\Theta = 0.1$, also equals (approximately) to the compressive strain at the inner edge ($\varepsilon_\text{in}=-\Theta$) and to the tensile strain at the outer edge ($\varepsilon_\text{out}=\Theta$).}
\label{fig:ribbons}
\end{figure}

Also supports are different, as the interaction with graphene can be either physical or chemical. In phy\-si\-sorption the support interaction is weak, graphene's electronic structure remains decoupled, and adhesion arises from the dispersive van der Waals interactions alone.\cite{abergel_AP_10} In chemisorption the support interaction is stronger, and the presence of chemical bonds alters graphene's electronic structure.\cite{olsen_PRL_11} Therefore adhesion, responsible for holding ribbons planar, ranges from $\epsilon =4$~meV/atom to $70$~meV/atom.\cite{olsen_PRL_11,koenig_nnano_11} However, fabrication processes, surface inhomogeneities, pinning, AFM tip manipulation, heat treatment, or mechanical strains can make ribbons subject to gentle bends, as sketched in Fig.~\ref{fig:ribbons}(a). Indeed, planar and gentle bending can be directly seen in scanning tunneling microscopy experiments.\cite{wang_nnano_11,li_science_08,jiao_nature_09,jiao_nnano_10} Distortions like twisting, on the contrary, are less relevant on supports.\cite{bets_NR_09,koskinen_APL_11,kit_PRB_12} Only gentle bends are interesting, as sharp bends are structurally unstable (ribbons would desorb and fold instead).\cite{li_science_08,jiao_nnano_10}


The purpose of this work, therefore, is to answer the following simple question: What happens to GNRs' electronic structure upon planar bending? It turns out that simple geometrical arguments, together with nearest (and next-nearest) neighbor tight-binding reasoning, are sufficient for a thorough understanding of the electromechanics of bent GNRs. These insights should hence help interpreting imperfect experiments with these distortion-prone ribbons.

\section{Simulating physisorbed ribbons with pure bending}

I modeled GNRs as free-standing, without explicit presence of the support; it was there merely as a planar constraint. The underlying justification was to model physisorption where the support and GNR electronic structures are essentially decoupled. For chemisorption the results are not directly valid. 

The focus is on the bent sections of very long ribbons, with bending viewed as a \emph{local} property. Apart from bending, planar ribbons can also stretch and shear; those deformations have been investigated by conventional methods.\cite{poetchke_PRB_10,wang_AA_12} The deformation mode certainly depends on the experimental conditions, and especially in short ribbons the strain patterns can become complicated.\cite{ribbon_note,young_AN_11} However, ribbons yield easily upon lateral forcing, and adjust themselves readily to minimum-energy geometries.\cite{bonelli_EPJB_09,lu_JAP_10,lee_science_10,cahangirov_PRL_12} In bent geometries sliding is particularly easy as the ribbon and the support are mostly out of registry. Long ribbons pinned at two distant locations, therefore, can remove high-energy shearing and stretching by sliding, and favor pure bending.\cite{pinning} I remark that the central results, as discussed below, will be valid also beyond pure bending.
  

I modeled the electronic structure by density-functional tight-binding (DFTB) method\cite{porezag_PRB_95,koskinen_CMS_09}, and the bent geometry itself by revised periodic boundary conditions.\cite{koskinen_PRL_10,dumitrica_JMPS_07,kit_PRB_11} Atoms in the simulation cell were from GNR translational cell of length $L$, and the associated symmetry operation was a rotation of an angle $\alpha$ around a given origin, as in Refs.~\onlinecite{dumitrica_JMPS_07} and \onlinecite{koskinen_PRB_10} [see Fig.~\ref{fig:ribbons}(b)]. This means, therefore, that the simulated systems were effectively GNR hoops, containing hundreds of thousands of atoms. Since the bends are gentle and the charge transfer with physisorption usually small, it's reasonable to assume that simulation describes the properties of bends in GNRs in a local sense.\cite{suenaga_nature_10} At any rate, regarding the bending, simulations were \emph{exact} and the sole approximation was the DFTB method itself.

Throughout this article I will use the dimensionless parameter
\begin{equation}
 \Theta = \frac{W}{2R},
\label{eq:theta}
\end{equation}
to quantify the amount of bending.\cite{malola_PRB_08b} Then, to simulate GNR of width $W$ with bending close to $\Theta'$, I chose $\alpha=L/R'$, with $R'=W/(2\Theta')$ as the initial guess for the radius of curvature, and optimized the structure. The only fixed parameter was $\alpha$, so $R$ and $\Theta$ were outcomes of the optimization, although $R\approx R'$ and $\Theta\approx \Theta'$. Here I remark that, because $\alpha$'s are small (down to $10^{-3}$ radians), the optimization was arduous and required maximum force criteria as small as $f_\text{max}<10^{-5}$~eV/\AA\ (look Ref.~\onlinecite{kit_PRB_11} to see why).

I conducted such simulations for hydrogen-passivated armchair ribbons ($N$-AGNRs) and zigzag ribbons ($N$-ZGNRs), with $N=5 \ldots 40$, with $W$ up to $84$~\AA, with $74$ different GNRs in total (see Ref.~\onlinecite{son_PRL_06} for GNR notations). Each GNR was optimized for $10$ bendings between $\Theta=0$ and $\Theta=0.1$. The reason for $\Theta=0.1$ as the upper limit for bending that I term ``gentle'' will be clarified later. Finally, the number of $\kappa$-points was $50$~\AA$/L$ for geometry optimization and $500$~\AA$/L$ for electronic structure analysis.


\begin{figure}
\includegraphics[width=8cm]{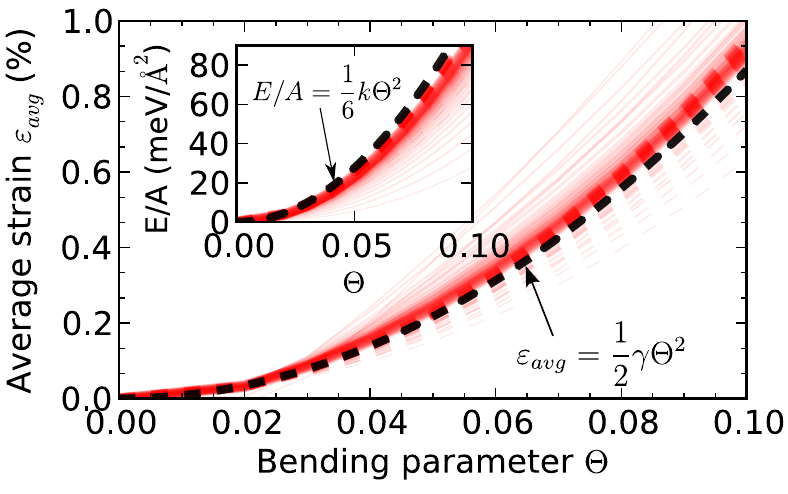}
\caption{(Color online) The cross-ribbon averaged strain $\varepsilon_\text{avg}$ as a function of bending parameter $\Theta$ for AGNRs (solid lines) and for ZGNRs (dashed lines); the bold dashed line is an analytical estimate. Inset: Elastic energy density as a function of bending; the bold dashed line is an analytical estimate, Eq.(\ref{eq:eavg}). In both plots the line width is proportional to $W$. Hence both in $\varepsilon_\text{avg}$ and in $E/A$ the largest deviations are for the narrowest ribbons.}
\label{fig:energy}
\end{figure}

\section{Bent ribbons get stretched}

Let us now turn attention to the results. Prior to discussing electronic properties, however, let us first look at energy and geometry. The energy in bent ribbon, as hinted by nanoshell elasticity\cite{kudin_PRB_01}, comes chiefly from axial in-plain strain. A quick estimate yields energy per unit area as $E/A=\frac{1}{6} k \Theta^2$, where $k=25$~eV\AA$^{-2}$ is graphene's in-plane modulus.\cite{landau_lifshitz} This simple estimate is in fair agreement with the simulations, as shown in the inset of Fig.~\ref{fig:energy}. Only the narrowest ribbons deviate from this estimate, for two reasons: First, the comparison of $W$ between atomic and continuum methods is inherently ambiguous; for small $W$ this ambiguity is emphasized. Second, in narrow ribbons the value of $k$ is affected by distinct elastic properties near the edges. While I could remedy these deficiencies by improving the model, my main interest is not in the minutiae of narrow ribbons, but in the wider ribbons and their universal trends.

If we set the adhesion energy $\epsilon$ equal to the strain energy, we get
\begin{equation}
 \Theta_\epsilon=\sqrt{6\epsilon/kA_c}
\end{equation}
as a rough estimate for the limit where the ribbon rather desorbs and straightens than remains adsorbed and bent on the support. The maximum adhesion $\epsilon=70$~meV/atom yields $\Theta_\epsilon=0.08$, justifying the upper limit $\Theta \lesssim 0.1$ for a ``gently`` bent ribbon, even though $\Theta_\epsilon$ really depends on the substrate. This is only an order-of-magnitude estimate, as fluctuations and finite-size effects can cause ribbons to desorb earlier. Direct experimental evidence\cite{li_science_08} shows how GNRs on SiO$_2$ bend up to $\Theta \approx 0.01$---still nearly half the simple-minded limit of $\Theta_\epsilon \approx 0.025$ given by $\epsilon \approx 6$~meV/\AA$^2$.\cite{miwa_APL_11}

Given the definition for $\Theta$, the strain on ribbon's inner edge is $\varepsilon_\text{in}\approx-\Theta$ and on the outer edge $\varepsilon_\text{out}\approx\Theta$. With strains around $10$~\%\ the bond anharmonicities begin to emerge, and stretching becomes cheaper, compression more expensive. This implies that the neutral line moves away from the origin ($R$ increases) and ribbon stretches. We can take this effect into account by a strain-dependent in-plane modulus, $k(\varepsilon)=k_0(1-\gamma \varepsilon)$. Then, by minimizing the total energy per unit length
\begin{equation}
 \int_{R-W/2}^{R+W/2} \frac{1}{2} k_0 (1-\gamma \varepsilon) \varepsilon^2 \der r
\end{equation}
with respect to $R$, we obtain the cross-ribbon averaged strain as
\begin{equation}
\varepsilon_\text{avg}=\frac{1}{2} \gamma \Theta^2.
\label{eq:eavg}
\end{equation}
This analytical estimate, given the value $\gamma=1.7$ obtained from DFTB simulations of stretched GNRs, agrees well with simulations, as shown in Fig.~\ref{fig:energy}. The largest deviations occur again for the narrow ribbons, albeit with opposite tendencies for AGNRs and ZGNRs due to different edge morphologies.

\begin{figure}
\includegraphics[width=8cm]{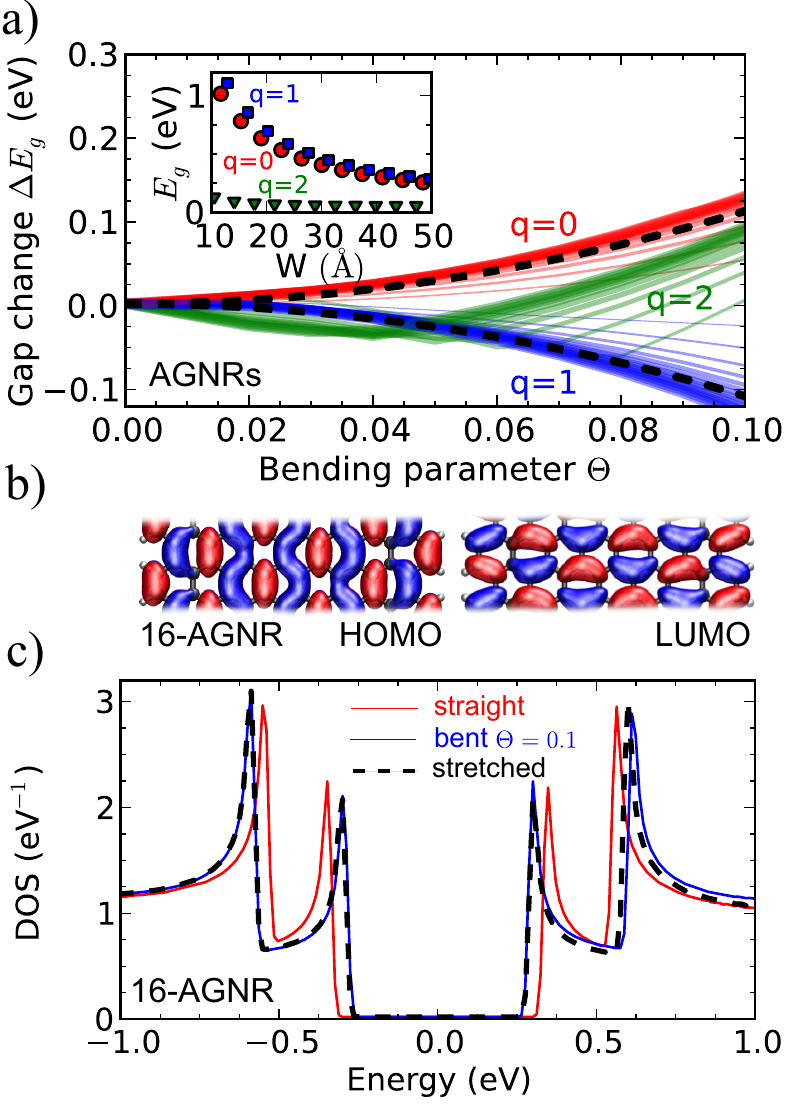}
\caption{(Color online) Electronic structure of bent AGNRs. (a) Bending-induced gap changes $E_g(\Theta)-E_g(0)$ for the three $q$-families of AGNRs. Line width is proportional to $W$; dashed lines are estimates for $q=0,1$. Inset: Gaps in straight AGNRs. (b) Wave functions for the frontier orbitals in straight $16$-AGNR ($q=1$): the highest occupied molecular orbital (HOMO) and the lowest unoccupied molecular orbital (LUMO). (c) Density of states for $16$-AGNR with straight ($\Theta=0.0$), bent ($\Theta=0.1$), and stretched ($\varepsilon=\frac{1}{2}\gamma (0.1)^2=0.85$~\%) geometries.}
\label{fig:ac}
\end{figure}

\section{Armchair ribbons are dominated by stretching}

Equipped with these geometrical notions, let us now turn attention to the electronic properties, starting with AGNRs. The inset in Fig.~\ref{fig:ac}(a) shows the energy gaps for the known three families of $N$-AGNRs, defined by $q=\mod (N,3)$.\cite{son_PRL_06,barone_NL_06} The gaps scale as $E_g\approx \beta W^{-1}$, where $\beta\approx 13$~eV\AA\ for $q=0,1$ (for $q=2$ the scaling is bit different). Figure~\ref{fig:ac}(a) shows how these gaps respond to bending: they widen or narrow with the same $q$-dependent families. Deviations occur only for the narrowest ribbons.


These trends can be understood by the following model. The energy gaps in stretched $q=0,1$ AGNRs depend on the strain as $\Delta E_g^\text{straight} \approx (-1)^q\varepsilon \delta$ with $\delta=12\text{ eV}$ (fit for $q=2$ is just more complex).\cite{hod_NL_09,poetchke_PRB_10} The origin for this strain-dependence is illustrated in Fig.\ref{fig:ac}(b) for $16$-AGNR with $q=1$: the highest occupied orbital is bonding and the lowest unoccupied orbital is antibonding along the ribbon's axis, and therefore stretching tends to narrow the gap (for $q=0$ AGNRs the situation is the opposite and for $q=2$ intermediate).\cite{gunlycke_NL_10} Next, if we pretend, in effect, that the bent AGNRs experience only the \emph{average} axial strain (even if the strain is uneven), and thus juxtapose $\varepsilon$ with $\varepsilon_\text{avg}$ from Eq.(\ref{eq:eavg}), we get
\begin{equation}
\Delta E_g (\Theta) \approx \frac{1}{2}(-1)^q \gamma \delta \Theta^2.
\end{equation}
Figure~\ref{fig:ac}(a) plots these estimates for $q=0$ and $q=1$ AGNRs by the dashed lines. The fair agreement suggests that the electronic structure of AGNRs subject to bending is \emph{dominated by the cross-ribbon averaged strain}. Similar physics has been observed previously in bent carbon nanotubes and twisted GNRs.\cite{koskinen_PRB_10,gunlycke_NL_10,koskinen_APL_11,zhang_small_11} Hence the argument is easily generalized to combined bending and stretching, where the electronic structure is modified by the average strain $\varepsilon_\text{stretch}+ \frac{1}{2} \gamma \Theta^2$.\cite{koskinen_APL_11} 

These trends, as given by four-valence DFTB, are reproduced by a $\pi$-only tight-binding Hamiltonian
\begin{equation}
H=-t\sum_{i,j}^\text{n.n.} c_i^\dagger c_j - t'\sum_{i,j}^\text{next n.n} c_i^\dagger c_j,
\label{eq:H}
\end{equation}
with the nearest-neighbor (n.n.) hopping parameter
\begin{equation}
t(r) = 2.6\text{ eV} - 5.8\text{ eV/\AA}(r-1.42\text{ \AA})
\label{eq:t}
\end{equation}
and with the next-nearest neighbor hopping equal to zero ($t'=0$, not shown). Figure~\ref{fig:ac}(c) shows further that the stretching analogy extends beyond energy gaps, as the entire density of states (DOS) is well described by the stretched geometry. This carries the average strain analogy also for optical transitions, as shown earlier.\cite{koskinen_PRB_10} Note that, if gauged through the relative gap change $|\Delta E_g/E_g| \approx 0.8\text{\AA}^{-1}W \Theta^2$, the influence of bending becomes more important as $W$ increases.


\section{Zigzag ribbons are dominated by broken symmetry}

\begin{figure}
\includegraphics[width=8cm]{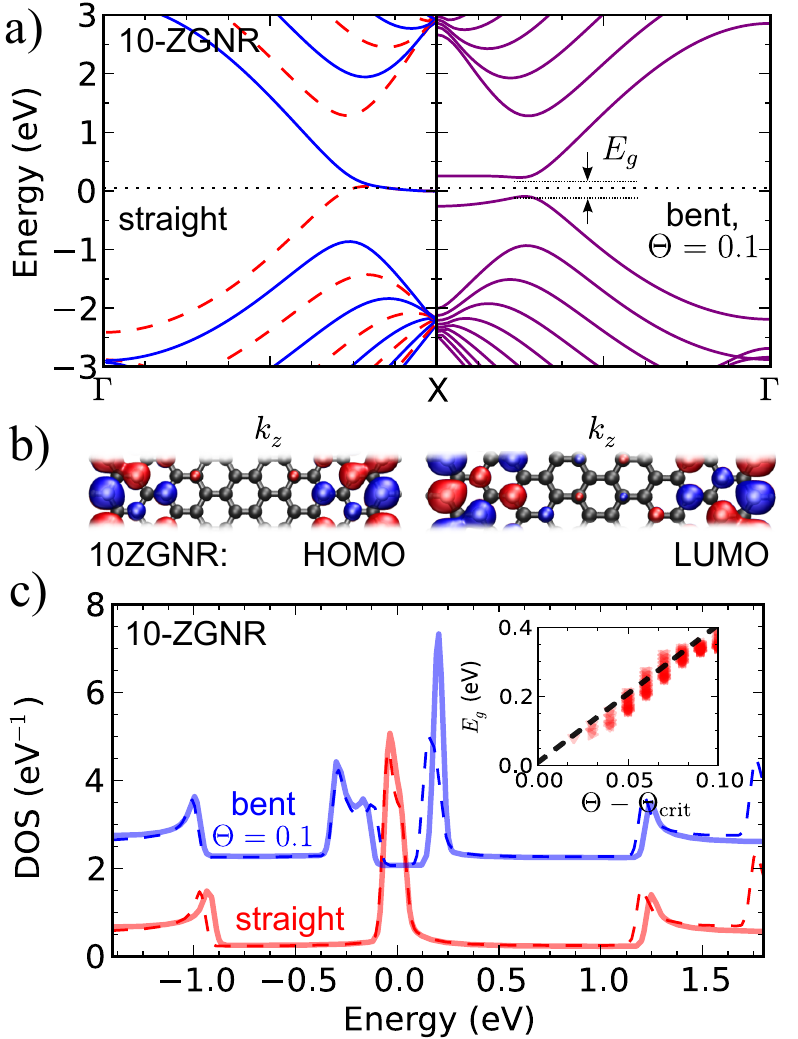}
\caption{(Color online) Electronic structure of bent ZGNRs. (a) Band structure of $10$-ZNGR with a straight and bent geometries. For the straight ribbon, having a reflection symmetry, the dashed lines denote symmetric and solid lines denote antisymmetric states under reflection. (For the bent ribbons no such distinction can be made.) (b) Wave functions of the frontier orbitals in straight $10$-ZGNR: highest occupied molecular orbital (HOMO, symmetric) and lowest unoccupied molecular orbital (LUMO, antisymmetric). (c) Density of states in straight and bent $10$-ZGNR. The dashed lines are from a next-nearest neighbor tight-binding model [Eq.(\ref{eq:H}) with Eq.(\ref{eq:tp})]. Inset: Gaps for all ZGNRs plotted as a function of $\Theta-\Theta_\text{crit}$ ($E_g=0$ when $\Theta<\Theta_\text{crit}$).}
\label{fig:zz}
\end{figure}

Let us now leave AGNRs and turn our attention to the electronic properties of ZGNRs. First I have to remind that, for straight ZGNRs as such, the spin-parallel DFTB simulations are dubious, given the prediction for a spin-polarized ground state.\cite{son_PRL_06} The magnetic structure should arise when the curious flat bands near the Fermi-level\cite{nakada_PRB_96} [left panel of Fig.\ref{fig:zz}(a)]---the famous edge states---spin-polarize, lift degeneracies, and open a gap (not shown).\cite{son_PRL_06} This way spontaneous magnetization can stabilize the electronic structure.


It has been shown that, unlike in AGNRs\cite{hod_NL_09}, the electronic structure in ZGNRs is unaffected by stretching.\cite{sun_JCP_08,poetchke_PRB_10} Therefore, after discovering above the average-strain argument with AGNRs, it's natural to guess that bending would leave ZGNRs' electronic structure unaffected. The right panel of Fig.~\ref{fig:zz}(a) shows, however, that \emph{bending can open an energy gap in ZGNRs}. That is, electronic structure is stabilized by sheer bending, and the cause for magnetic spin-polarization is lost. Note how the band structure changes only near the Fermi-level, while other bands remain stable.

The mechanism of the gap opening is related to broken reflection symmetry, as clarified by the following three-step reasoning: First step; the bands are called ''flat'' because they have small dispersion. In the absence of next-nearest neighbor hopping ($t'=0$), the Hamiltonian (\ref{eq:H}) gives edge states whose dispersion and energy are essentially zero, independent of $t$. This is illustrated in Fig.~\ref{fig:zz}(b), where flat band electrons appear localized to next-nearest neighbor sites, separated by vacancies. Therefore the $t$ in Eq.~(\ref{eq:t}), even if strain-dependent, doesn't affect the flat bands---splitting and dispersion hence require next-nearest neighbor hopping $t'$. Second step; fitting $t'$ to strained GNRs by DFTB gives
\begin{equation}
t'(r)= 0.25 \text{ eV} -0.6\text{ eV\AA}(r-2.46\text{ \AA}).
\label{eq:tp}
\end{equation}
The Hamiltonian (\ref{eq:H}), with hoppings (\ref{eq:t}) and (\ref{eq:tp}), reproduces the electronic structure fairly well, as shown by the DOS for $10$-ZGNR in Fig.~\ref{fig:zz}(c). [Pure stretching leaves DOS intact (not shown)]. Third step; the flat band energies (also the band dispersion) are proportional to $t'$ and hence proportional to edge strain via Eq.(\ref{eq:tp}). Upon bending, the reflection symmetry breaks and states localize on either of the edges with strain difference $\varepsilon_\text{out}-\varepsilon_\text{in}=2\Theta$; opposite edges hence get unequal hoppings $\Delta t'=t'_\text{out}-t'_\text{in} \propto \Theta$. Because energy splitting is proportional to $\Delta t'$, it is also proportional to $\Theta$. This is the mechanism how bending splits the flat bands with direct proportionality to $\Theta$.

As mentioned above, since $t'$ gives flat bands a small dispersion, splitting does not open the gap immediately. When $W$ increases, the span of the flat region in $k_z$-space increases, and gap opening requires larger splitting. A fit to all ZGNRs yields a critical value for opening a gap as $\Theta_\text{crit}\approx W/200$~nm (or $R_\text{crit}\approx 100$~nm for all $W$), yielding the energy gap as
\begin{equation}
 E_g^\text{ZGNR} \approx 4 \text{ eV} (\Theta - \Theta_\text{crit}).
\label{eq:zzgap}
\end{equation}
The gaps, displaying values up to $0.4$~eV, are plotted in the inset of Fig.~\ref{fig:zz}(c).


\section{Conclusions}

The physics in AGNRs and ZGNRs appear hence quite different: AGNRs are governed by average strain, whereas ZGNRs are governed by broken reflection symmetry. The effects of broken symmetry on AGNRs or average strain on ZGNRs surely exist, but they are just less important. In ZGNRs bending can have particular impact on transport, since the localization of edge states depends on the direction of bending; if ribbon has bends both to the left and to the right, the current-carrying electrons need to jump from one edge to the other, suggesting width-dependent resistivity.\cite{han_PRL_10} Although it's plausible that bent ZGNRs indeed acquire gaps and turn nonmagnetic, spin-polarized calculations would be opportune, even if the existence of magnetism has been disputed also for the straight ribbons.\cite{kunstmann_PRB_11}

I obtained similar results also for unpassivated GNRs, observing similar phenomena. Thus it appears that these clear trends arise from simple physics with plausible explanations, and it's unlikely that, say, higher level electronic structure methods should change the picture. I believe, therefore, that these general trends are helpful enough to serve as rules of thumb to aid GNR device fabrication and analysis.

\emph{Acknowledgements} I acknowledge Teemu Peltonen for discussions, the Academy of Finland for funding, and the Finnish IT Center for Science (CSC) for computer resources.


\begin{thebibliography}{44}
\bibitem[{rib()}]{ribbon_note}
\bibinfo{note}{Short means that ribbon's length is of the same order of magnitude than its width.}
\end{thebibliography}

\end{document}